\documentclass[prd,aps,preprint,showpacs,nofootinbib,eqsecnum,superscriptaddress]{revtex4-1}

\usepackage{latexsym}
\usepackage{hyperref}
\usepackage{epstopdf}
\usepackage{amsfonts,amssymb,amsmath,amsthm,amsxtra,mathtools,euscript,mathrsfs}
\usepackage{dsfont}
\usepackage{yhmath}
\usepackage{arcs}
\usepackage{slashed}
\usepackage{cancel}
\usepackage{epsf}
\usepackage{epsfig}

\usepackage{bm}
\usepackage{graphicx,color,xcolor,subfigure}
\usepackage[font=small,labelfont=bf]{caption}
\usepackage{dcolumn}
\usepackage{array}
\usepackage{tabularx}
\usepackage{enumerate}
\usepackage{hhline}
\usepackage{xfrac}

\def\a{\alpha}
\def\b{\beta}

\def\g{\gamma}
\def\s{\sigma}
\def\k{\kappa}

\def\vt{\vartheta}

\def\veps{\varepsilon}

\def\k{\kappa}
\def\k5{\kappa_5}

\newcommand{\ie}{\textit{i}.\textit{e}.}

\newcommand{\lagrangian}{\mathcal{L}}

\def\hzero{\hat{0}}
\def\hone{\hat{1}}
\def\htwo{\hat{2}}
\def\hthree{\hat{3}}

\def\hfive{\hat{5}}

\begin{document}
\title{Kaluza-Klein theory for teleparallel gravity}

\author{Chao-Qiang Geng}
\email[Electronic address: ]{geng@phys.nthu.edu.tw}
\affiliation{Chongqing University
of Posts \& Telecommunications, Chongqing, 400065, China}
\affiliation{Department of Physics,
National Tsing Hua University, Hsinchu 300, Taiwan}
\affiliation{Physics Division,
National Center for Theoretical Sciences, Hsinchu 300, Taiwan}
\author{Chang Lai}
\email[Electronic address: ]{laichang@cqupt.edu.cn}
\affiliation{Chongqing University
of Posts \& Telecommunications, Chongqing, 400065, China}
\author{Ling-Wei Luo}
\email[Electronic address: ]{d9622508@oz.nthu.edu.tw}
\affiliation{Department of Physics,
National Tsing Hua University, Hsinchu 300, Taiwan}
\author{Huan-Hsin Tseng}
\email[Electronic address: ]{d943335@oz.nthu.edu.tw}
\affiliation{Department of Physics,
National Tsing Hua University, Hsinchu 300, Taiwan}

\begin{abstract}
We study teleparallel gravity in the \emph{original} Kaluza-Klein (KK) scenario.
Our calculation of the KK reduction of teleparallel gravity indicates that the 5-dimensional
torsion scalar $^{(5)}T$ generates the non-Brans-Dicke type effective Lagrangian in 4-dimension
due to an additional coupling between the derivative of the scalar field and torsion,
but the result is  equivalent to that in general relativity. We also discuss the cosmological
behavior in the FLRW universe based on the effective teleparallel gravity.
\end{abstract}


\maketitle

\begin{section}{Introduction}

The extra dimension theory was originally proposed to unify the electromagnetism
and gravity theory into one theory via 5-dimension with gauge invariant suggested by Kaluza~\cite{Kaluza}.
Klein then realized the Kaluza's cylindrical condition as the \emph{zero mode} of harmonic expansion fields
with a compactification of the 5th-dimension into a small circle $S^1$~\cite{Klein}.
In the Kaluza-Klein (KK) theory, particles are described by series of the mass spectrum, called as the KK towels.
The contributions from the extra dimension  give rise to the different mass scales by the KK towels.
In the low energy scale, the KK dimensional reduction leads to the effective theory with gravity
interacting with electromagnetic and scalar fields in 4-dimension~\cite{Overduin:1998pn}.

An alternative gravity theory with \emph{absolutely parallelism}
in \emph{Weitzenb\"{o}ck geometry} called \emph{teleparallel equivalent to general relativity} (TEGR)
was first considered by Einstein~\cite{Einstein:ap}.
Recently, several generalizations related to \emph{teleparallel gravity} (teleparallelism) have been
presented in the literature, such as teleparallel dark energy~\cite{Geng11} and
$f(T)$~\cite{Ferraro:2006jd,Linder:2010py} models.

Teleparallelism in the KK scenario has been
discussed in Refs.~\cite{deAndrade:1999vq, Barbosa:2002mg,
Fiorini:2013hva, Bamba:2013fta, Geng:2014yya}. In this article,
we  compute the KK reduction of TEGR at the low energy
in the absence of the electromagnetic field.
The reduction generates an additional non-minimal
coupled term for the effective action,
which  matches 
neither with the result in Ref.~\cite{Bamba:2013fta} nor with
the Brans-Dicke type theory.
We also show that there exists an \emph{Einstein frame} for
the non-minimal teleparallel gravity by including the additional coupling.
Finally, we study the flat Friedmann-Lema\^{i}tre-Robertson-Walker (FLRW)
cosmology and discuss the solution in the KK scenario.

\end{section}

\begin{section}{Kaluza-Klein Reduction for Teleparallel Gravity}

\begin{subsection}{Brief Review of Teleparallel Gravity}
In the teleparallel theory, one introduces an \emph{orthonormal frame $\vartheta^i = e^i_{\mu} \,dx^{\mu}$
with a given metric $g_{\mu\nu}=\eta_{ij}\, e^i_{\mu}\, e^j_{\nu}$,
where $\eta_{ij}=\text{diag}(+1,-1,-1,-1)$ is Minkowski metric,
$\mu,\nu,\rho,\ldots=0,1,2,3$ label coordinate frame indices,
and $i,j,k,\ldots = \hzero,\hone,\htwo,\hthree$ denote orthonormal frame indices.
As a result, we have $e_i=e^{\mu}_i\,\partial_{\mu}$ and $\vt^i = e^{i}_{\mu}\,dx^{\mu}$
as the fields in the orthonormal frame, also called \emph{vielbein} fields denoted by
the coefficients $e^i_\mu$. The duality of coframes and frames leads to the relations
$e^i_{\mu}\,e^{\mu}_j=\delta^i_j$ and $e^{\mu}_i\,e^i_{\nu}=\delta^{\mu}_{\nu}$.}
One can define the \emph{absolute parallelism}, \ie, the \emph{curvature-free} condition, via
the covariant derivative of vielbein fields of
$\nabla_{\nu}e^i_\mu=\partial_{\nu}e^i_\mu-e^i_{\rho}\, \Gamma^{\rho}_{\mu\nu}=0$
to obtain the \emph{Weitzenb\"{o}ck connection}
$\Gamma^{\rho}_{\mu\nu}= e^{\rho}_i\partial_{\nu}e^i_{\mu}$ in Weitzenb\"{o}ck geometry.
The torsion tensor is defined by the anti-symmetric part of the Weitzenb\"{o}ck connection
$T^{\rho}{}_{\mu\nu}\equiv \Gamma^{\rho}_{\nu\mu}-\Gamma^{\rho}_{\mu\nu}=
e^{\rho}_i(\partial_{\mu}e^i_{\nu}-\partial_{\nu}e^i_{\mu})= -T^{\rho}{}_{\nu\mu}$
and the \emph{contorsion}
$K^{\rho}{}_{\mu\nu}\equiv (-1/2)(T^{\rho}{}_{\mu\nu} -
T_{\mu}{}^{\rho}{}_{\nu} - T_{\nu}{}^{\rho}{}_{\mu}) = -K_{\mu}{}^{\rho}{}_{\nu}$.
The \emph{torsion scalar} is constructed as~\cite{Einstein:teleaction}
\begin{subequations}\label{E:torsion scalar}
\begin{eqnarray}
T &=& \frac{1}{4}\,T_{\rho\mu\nu} \, T^{\rho\mu\nu} +
      \frac{1}{2}\,T_{\rho\mu\nu} \,T^{\nu\mu\rho} -
      \,T^{\nu}{}_{\mu\nu} \, T^{\s\mu}{}_{\s}\\
  &=& \frac{1}{4}\,T_{kij} \, T^{kij} +
      \frac{1}{2}\,T_{kij} \,T^{jik} -
      \,T^{j}{}_{ij} \, T^{ki}{}_{k}\,.\label{E:torsion scalar in orthonormal frame}
\end{eqnarray}
\end{subequations}
The torsion scalar can also be rewritten as
$T \equiv \frac{1}{2}\,T^{\rho}{}_{\mu\nu}\,S_{\rho}{}^{\mu\nu}$,
where $S_{\rho}{}^{\mu\nu}\equiv K^{\mu\nu}{}_{\rho} +
\delta_{\rho}^{\mu}\,T^{\s\nu}{}_{\s} -
\delta_{\rho}^{\nu}\,T^{\s\mu}{}_{\s} =
-S_{\rho}{}^{\nu\mu}$
is called the \emph{superpotential}.
In general relativity, the Hilbert Lagrangian is the curvature scalar $R$.
In analogy, the teleparallel gravitational
action of TEGR is given by the torsion scalar
\begin{equation}\label{E:telaparallel Lagrangian}
S_{\text{TEGR}}= \frac{1}{2\kappa} \int d^4x\, e\, T\,,
\end{equation}
where $\kappa= 8\pi\,G$ is the gravitational coupling and
$e\equiv \det(e^i_{\mu})$.

In the 5-dimensional teleparallel gravity, the metric is given by
$\bar{g}_{MN}=\bar{\eta}_{IJ}\, e^I_{M}\, e^J_{N}$,
where $\bar{\eta}_{IJ}=\text{diag}(+1,-1,-1,-1,\veps)$
with $\veps=\pm1$, $M,N,O=0,1,2,3,5$ and $I,J,K=\hzero,\hone,\htwo,\hthree,\hfive$.
It is convenient to calculate torsion by the \emph{Cartan structure equation} $T^{I}=d\theta^{I}$.
The non-vanishing components of the torsion 2-form in the orthonormal frame are
$\bar{T}^{k}{}_{ij}, \bar{T}^{k}{}_{5j}$
and $\bar{T}^{5}{}_{i5}$.
The 5-dimensional torsion scalar ${}^{(5)}T$ can be given in a similar way as
Eq.~(\ref{E:torsion scalar in orthonormal frame}) in 5-dimension,
decomposed as~\cite{Geng:2014yya}
\begin{equation}
{}^{(5)}T= \bar{T} + \frac{1}{2}\left(
\bar{T}_{k5j}\,\bar{T}^{k5j} +
\bar{T}_{k5j}\,\bar{T}^{j5k}\right) +
2\,\bar{T}^{k}{}_{k}{}^{i}\,\bar{T}^{5}{}_{i5} -
\bar{T}^{j}{}_{5j}\,\bar{T}^{k5}{}_{k}\,,
\end{equation}
where $\bar{T}$ is the induced 4-dimensional torsion scalar.

\end{subsection}

\begin{subsection}{Kaluza-Klein Theory}
In the original Kaluza-Klein theory, the 5-dimensional spacetime
bulk $V=M\times S^1$ is a \emph{product space} of a hypersurface $M$ and
a circle $S^1$ with small radius $r$ representing the
\emph{compactified extra dimension}.
Because of the embedding, we can always write
the 5-dimensional metric $\bar{g}$ in the coordinate system as
\begin{equation}
\bar{g}_{MN} =
\begin{pmatrix}
  g_{\mu\nu}(x^{\mu},y)  &  0  \\
  0                      &  \veps \phi^2(x^{\mu},y)
\end{pmatrix}\,,
\end{equation}
where $y=x^5$ and $\veps=-1$.
In the orthonormal frame, the vielbein fields are $e^i_{\mu}$ and
$e^{\hfive}_{5}=\phi(x^{\mu},y)$.
Consequently, the 5-dimensional torsion scalar in the coordinate frame is
\begin{equation}
{}^{(5)}T= \bar{T} + \frac{1}{2}\left(
\bar{T}_{\rho 5\nu}\,\bar{T}^{\rho 5\nu} +
\bar{T}_{\rho 5\nu}\,\bar{T}^{\nu 5\rho}\right) +
2\,\bar{T}^{\s}{}_{\s}{}^{\mu}\,\bar{T}^{5}{}_{\mu5} -
\bar{T}^{\nu}{}_{5\nu}\,\bar{T}^{\s5}{}_{\s}\,,
\end{equation}
where $\bar{T}=T$ is a pure 4-dimensional object
due to $\bar{T}^{\rho}{}_{\mu\nu}= T^{\rho}{}_{\mu\nu}$.
Note that as there only exists the KK zero mode in the \emph{effective low-energy theory}
based on the \emph{cylindrical condition},
all fields have no dependence on the 5th-dimensional component,
\ie, $g_{\mu\nu,5} =0 $, which is referred to as
\emph{KK ansatz}~\cite{Overduin:1998pn}.
Within the ansatz, the metric is reduced to
\begin{equation}\label{E:KK metric}
\bar{g}_{MN} =
\begin{pmatrix}
    g_{\mu\nu}(x^{\mu})  &  0  \\
    0                    &  -\phi^2(x^{\mu})
\end{pmatrix}\,.
\end{equation}
The residual components of the non-vanishing torsion tensor are
$T^{\rho}{}_{\mu\nu}$ and $\bar{T}^{5}{}_{\mu5}=\partial_{\mu}\phi/\phi$.
The 5-dimensional torsion scalar is
\begin{equation}\label{E:KK torsion sclar}
{}^{(5)}T= T + 2\,T^{\s}{}_{\s}{}^{\mu}\,\bar{T}^{5}{}_{\mu5}\,,
\end{equation}
which can be used to construct
the 5-dimensional action.
The extra dimension $y=r\,\theta$ is the ground state of $\bar{g}_{55}$ with the radius
$r$.
The invariant volume element is ${}^{(5)}e\,d^5x=e\,\phi\,r\,d\theta\,d^4x$
by the dimensional reduction with ${}^{(5)}e=\det(e^I_{M})$.
As a result, we obtain the 4-dimensional effective action
\begin{equation}\label{E:Effective action of T}
S_{\text{eff}}= \frac{1}{2\,\kappa_4} \int d^4x\, e
\left( \phi\, T + 2\,T^{\mu}\,\partial_{\mu}\phi \right)\,,
\end{equation}
where the effective coupling $\kappa_4:= \k5/2\pi r= 8\pi\,G_5/2\pi r$ and
$T^{\mu}:=T^{\s}{}_{\s}{}^{\mu}$ is the torsion trace vector.
Hence, we have the KK reduction procedure in teleparallel gravity which
replaces the 5-dimensional torsion scalar ${}^{(5)}T$
instead of $T + 2\,{\phi}^{-1}\,T^{\mu}\,\partial_{\mu}\phi$.

For $F(T)$ gravity in the KK theory, where $F(T)$ is an arbitrary function of $T$,
the Lagrangian density ${}^{(5)}e\,F({}^{(5)}T)$ becomes
$2\pi\,e\,\phi\,F(T + 2\,{\phi}^{-1}\,T^{\mu}\,\partial_{\mu}\phi)$.
The effective action for $F(T)$ gravity is
\begin{equation}
\label{E:Effective action of F(T)}
{\bf S}_{\text{eff}}= \frac{1}{2\,\kappa_4} \int d^4x\, e\,\phi\,
F\left( T + \frac{2}{\phi}\,T^{\mu}\,\partial_{\mu}\phi \right)\,.
\end{equation}
By taking $F({}^{(5)}T)$ to be linear in the torsion scalar with a cosmological constant,
\ie, $F({}^{(5)}T)= {}^{(5)}T-2\,\Lambda_5$, the action can be reduced to
Eq.~(\ref{E:Effective action of T}) with a \emph{varying} 4-dimensional
cosmological term $\Lambda_4(x^{\mu})= \Lambda_5\,\phi(x^{\mu})$.

It is interesting to consider a conformal transformation
$\tilde{g}_{\mu\nu}= \Omega^2(x^{\mu})\, g_{\mu\nu}$ for the
effective action in Eq.~(\ref{E:Effective action of T}).
The transformations of the vierbien and torsion tensor are
$\tilde{e}_{\mu}^{i}=\Omega\,e_{\mu}^{i}$ and
$\tilde{T}^{i}{}_{\mu\nu}=\Omega\,T^{i}{}_{\mu\nu} +
e_{\nu}^{i}\,\partial_{\mu}\Omega - e_{\mu}^{i}\,\partial_{\nu}\Omega$,
while the torsion scalar and the torsion trace vector are expressed as
\begin{subequations}
\begin{eqnarray}
T & = & \Omega^2\,\tilde{T} -
4\,\Omega\,\tilde{g}^{\mu\nu}\,\tilde{T}_{\mu}\,\partial_{\nu}\Omega -
6\,\tilde{g}^{\mu\nu}\,\partial_{\mu}\Omega\,\partial_{\nu}\Omega\,, \\
T_{\mu} & = & \tilde{T}_{\mu} + 3\,\Omega^{-1}\,\partial_{\mu}\Omega\,,
\end{eqnarray}
\end{subequations}
respectively. By choosing $\phi= \Omega^2$, the action reads
\begin{equation}\label{E:Conformal transformation of T}
S_{\text{eff}}= \int d^4x\, \tilde{e}\,\left[\frac{1}{2\,\kappa_4}\,\tilde{T}
+\frac{1}{2}\tilde{g}^{\mu\nu}\partial_{\mu}\psi\,\partial_{\nu}\psi\right]\,,
\end{equation}
where $\psi=(6/\kappa_4)^{1/2}\ln\Omega$
is a dilaton field.
As seen from Eq.~(\ref{E:Conformal transformation of T}),
there exists an \emph{Einstein frame}
for the non-minimal coupled effective Lagrangian in Eq.~(\ref{E:Effective action of T})
in teleparallel gravity.

By using Eq.~(\ref{E:Effective action of T})  and
the curvature-torsion relation
\begin{equation}\label{E:Curvature-Torsion Relation}
-\tilde{R}(e) = T - 2 \tilde{\nabla}_{\mu}T^{\mu}\,,
\end{equation}
where $\tilde{R}(e)$ and $\tilde{\nabla}_{\mu}$ are
the Riemann curvature constructed by
the \emph{Levi-Civita connection} in terms of the vielbein and
 the covariant derivative with respect to the Levi-Civita connection, respectively,
we obtain the action
\begin{equation}\label{E:Equivalent action of GR}
\frac{-1}{2\kappa_4}\int d^4x\,e\bigg(\phi\,\tilde{R}(e) -
2\,\tilde{\nabla}_{\mu}(\phi\,T^{\mu})\bigg)\,.
\end{equation}
It is interesting to note that in the 5-dimensional general relativity,
the effective action is $(-1/2\kappa_4)\int d^4x \sqrt{-g}\,\phi\,\tilde{R}$
which is a specific case of the Brans-Dicke theory~\cite{Overduin:1998pn,Fujii:2003pa,Carroll:2004st}.
It is clear that the form of the effective action in Eq.~(\ref{E:Effective action of T})
is not a scalar-tensor-like due to the additional non-minimal coupling $2\,T^{\mu}\partial_{\mu}\phi$, 
which is different from the reduction action in general relativity although it is still \emph{equivalent}
to the Lagrangian $\phi \tilde{R}$ up to the total derivative term
as shown in Eq.~(\ref{E:Equivalent action of GR}).
We also note that our action in Eq.~(\ref{E:Effective action of F(T)})
is clearly different from Eq.~(5) in Ref.~\cite{Bamba:2013fta}, in which the function variable
is $T + \phi^{-2}\,\partial^{\mu}\phi\,\partial_{\mu}\phi$ without  coupling between $T$ and $\phi$
instead of $T + 2\,{\phi}^{-1}\,T^{\mu}\,\partial_{\mu}\phi$.

\end{subsection}

\end{section}

\begin{section}{Equations of Motion}

By varying Eq.~(\ref{E:Effective action of T}) with respect to $e^{i}_{\mu}$,
the equation of motion of teleparallel gravity is
\begin{eqnarray}
\frac{1}{2}\,e^{\mu}_{i}\bigg(\phi\,T+2\,T^{\s}\,\partial_{\s}\phi\bigg) -
e^{\rho}_{i}\bigg(\phi\,T^{j}{}_{\rho\nu}\,S_{j}{}^{\mu\nu}\bigg)
& - &
e^{\nu}_{i}\bigg(\partial_{\s}\phi\,T^{\mu}{}_{\nu}{}^{\s} +
\partial_{\nu}\phi\,T^{\mu} + \partial^{\mu}\phi\,T_{\nu}\bigg)
\nonumber \\
+\,\frac{1}{e}\, \partial_{\nu}\bigg(e\,(\phi\,S_{i}{}^{\mu\nu} +
e^{\mu}_{i}\,\partial^{\nu}\phi -
e^{\nu}_{i}\,\partial^{\mu}\phi)\bigg)
& = &
\kappa_4\,\Theta^{\mu}_{i}\,,
\end{eqnarray}
with $\Theta^{\mu}_{i}=
(-1/e)(\delta\,\lagrangian_m/\delta\,e^{i}_{\mu})$, where
we have assumed the energy-momentum tensor as a perfect fluid with
 $\Theta^{\mu}_{\nu}=\text{diag}(\rho,-P,-P,-P)$. For the flat FLRW
universe, we have $g_{\mu\nu}=
\text{diag}(1,-a^2(t),-a^2(t),$$-a^2(t))$ and
 $e^{i}_{\mu}= \text{diag}(1,a(t),a(t),a(t))$.
In such coordinates, the non-vanishing torsion component is $T^{\a}{}_{0\a}=\dot{a}/a$
without summation, where $\a,\b,\g,\ldots=1,2,3$ are in the coordinate frame
and $a,b,c,\ldots=\hone,\htwo,\hthree$ in the orthonormal frame.
The only torsion trace vector of $T^{\mu}$ is the $\mu=0$ component, given by
$T^{0}= 3\,T^{\a}{}_{\a}{}^{0}= -3\,g^{00}\,T^{\a}{}_{0\a}= (-3)(\dot{a}/a)$.
From the identity $K^{\rho}{}_{\mu\rho}= T^{\rho}{}_{\rho\mu}$,
we have $K^{\a}{}_{0\a}= -\dot{a}/a$, $K^{0}{}_{\a\a}=a\,\dot{a}$, and
$S_{\a}{}^{0\a}= (-2)(\dot{a}/a)$.
The Friedmann equations are given by
\begin{subequations}\label{E:Friedmann equations}
\begin{eqnarray}
3\,\phi\,H^2+3\,H\,\dot{\phi} &=& \kappa_4\,\rho\,, \\
3\,\phi\,H^2 +2\,\dot{\phi}\,H +
2\,\phi\,\dot{H} + \ddot{\phi} &=& -\,\kappa_4\,P\,,
\end{eqnarray}
\end{subequations}
where $H=\dot{a}/a$ is the Hubble parameter.
It can be checked that Eq.~(\ref{E:Friedmann equations}) can be reduced to
the Friedmann equations of TEGR by taking $\phi=1$.

The equation of motion of the scalar field $\phi$ is
\begin{equation}
T - 2\,\partial_{\mu}T^{\mu}-2T^{\mu}\Gamma^{\nu}_{\nu\mu}= 0\,,
\end{equation}
where $\Gamma^{\nu}_{\nu\mu}=e^{\nu}_{i}\partial_{\mu}e^{i}_{\nu}$ 
with the non-vanishing component being $e^{\a}_{a}\partial_{0}e^{a}_{\a}=3\,(\dot{a}/a)$ 
in the coordinate. Subsequently, we obtain
$a\,\ddot{a}+\,\dot{a}^2=0$. The equation of motion is independent
of the scalar field and the scale factor can be solved directly. By
taking the solution to be proportional to 
$t^m$, 
we find that $m=0$ and $1/2$,
 leading to
\begin{equation}\label{E:Scale factor}
a(t) = a_s + b\,\sqrt{t},
\end{equation}
where $a_s$ ($>0$) and $b$ are constants of $\mathcal{O}(1)$. 
By substituting the solution in Eq.~(\ref{E:Scale factor}) back to $a\,\ddot{a}+\,\dot{a}^2=0$,
we find that the condition $a_s \cdot b=0$ must hold, which results in 
two cases: (i) $a_s\neq0$, $b=0$ and  (ii) $a_s=0$, $b\neq0$.

For (i), we obtain $a=a_s$, which corresponds to  a \emph{static} universe, where
$a_s$ is the scale factor of the valid energy scale
for the effective teleparallel gravity in Eq.~(\ref{E:Effective action of T}).
For (ii), we get $a=b\,\sqrt{t}$. In this case, 
$t$ has to be larger than $t_{\text{cut}}$ which represents the cut-off energy scale
for the low-energy effective teleparallel gravity.
Consequently, $\ddot{a}=-b/(4\,t^{2/3})$ and $H= 1/(2\,t)$.
For $b<0$ ($>0$), the universe is accelerated (decelerated) expanding.

Comparing to the effective Lagrangian in general relativity, 
the equation of motion of $\phi$ is $\tilde{R}(e)=0$, 
which also leads to the same solution for the scale factor.
 We conclude that  teleparallel gravity and 
general relativity are equivalent in the KK scenario.

Finally, we remark that the KK reduction generates an effective low-energy theory
that is improper to be applied  to the inflationary stage.
At the high energy scale, it has to consider the KK modes of
gravitational and scalar fields, \ie,
there exist massive gravitational and scalar fields.
Clearly, the situation is much more complicated than
the one discussed in Ref.~\cite{Bamba:2013fta}.

\end{section}

\begin{section}{Conclusions}

We have examined the KK reduction of telaparallel gravity.
Our result in Eq.~(\ref{E:Effective action of T}) has  shown that there is
a coupled term between the derivative of the scalar field and torsion trace vector,
which implies that the KK reduction
procedure can not be applied in telaparallelism to obtain the Brans-Dicke type theory.
The effective Lagrangian is different from general relativity
although it is equivalent to $\phi \tilde{R}$ up to the total
derivative term.
The property of the additional coupling leads to an Einstein frame
by the conformal transformation for the non-minimal coupled teleparallel gravity,
which is different from the result in the literature~\cite{Bamba:2013fta}.
In cosmology, we have obtained
the equation of motion in the FLRW universe and found that
the accelerated expansion of the universe can be achieved by the effective teleparallel gravity,
which is the same as the effective KK scenario in general relativity.

\end{section}

\begin{acknowledgments}
We thank Professor K.~Bamba for communications and discussions.
The work was supported in part by National Center for Theoretical Sciences, National Science
Council (NSC-101-2112-M-007-006-MY3) and National Tsing Hua
University (103N2724E1).
\end{acknowledgments}



\begin{thebibliography}{2}

\bibitem{Kaluza}
  T.~Kaluza,
  Sitzungsber.\ Preuss.\ Akad.\ Wiss.\ Berlin (Math.\ Phys.\ ) {\bf 1921}, 966 (1921).

\bibitem{Klein}
  O.~Klein,
  Z.\ Phys.\  {\bf 37}, 895 (1926)
  [Surveys High Energ.\ Phys.\  {\bf 5}, 241 (1986)].

\bibitem{Overduin:1998pn}
  J.~M.~Overduin and P.~S.~Wesson,
  Phys.\ Rept.\  {\bf 283}, 303 (1997).

\bibitem{Einstein:ap}
  A.~Einstein,
  Sitzungsber. Preuss. Akad. Wiss., {\bf 1928}(XVII), 217–221, (1928).

\bibitem{Geng11}
 C.~Q.~Geng, C.~C.~Lee, E.~N.~Saridakis and Y.~P.~Wu,
  Phys.\ Lett.\ B {\bf 704}, 384 (2011);
 C.~Q.~Geng, C.~C.~Lee and E.~N.~Saridakis,
  JCAP {\bf 1201}, 002 (2012);
    J.~A.~Gu, C.~C.~Lee and C.~Q.~Geng,
  Phys.\ Lett.\ B {\bf 718}, 722 (2013);
   C.~Q.~Geng, J.~A.~Gu and C.~C.~Lee,
  Phys.\ Rev.\ D {\bf 88}, 024030 (2013);
J.~T.~Li, Y.~P.~Wu and C.~Q.~Geng,
  Phys.\ Rev.\ D {\bf 89}, 044040 (2014).

\bibitem{Ferraro:2006jd}
  R.~Ferraro and F.~Fiorini,
  Phys.\ Rev.\ D {\bf 75}, 084031 (2007).

  \bibitem{Linder:2010py}
  E.~V.~Linder,
  Phys.\ Rev.\ D {\bf 81}, 127301 (2010)
  [Erratum-ibid.\ D {\bf 82}, 109902 (2010)];
  K.~Bamba, C.~Q.~Geng and C.~C.~Lee,
  arXiv:1008.4036 [astro-ph.CO];
  P.~Wu and H.~W.~Yu,
  Eur.\ Phys.\ J.\ C {\bf 71}, 1552 (2011);
  K.~Bamba, C.~Q.~Geng, C.~C.~Lee and L.~W.~Luo,
  JCAP {\bf 1101}, 021 (2011);
  S. H. Chen, J. B. Dent, S. Dutta and E. N. Saridakis,
  Phys. Rev. D {\bf 83}, 023508 (2011);
  Y. F. Cai, S. H. Chen, J. B. Dent, S. Dutta and E. N. Saridakis,
  Class. Quant. Grav.\  {\bf 28}, 215011 (2011).


\bibitem{deAndrade:1999vq}
  V.~C.~de Andrade, L.~C.~T.~Guillen and J.~G.~Pereira,
  Phys.\ Rev.\ D {\bf 61}, 084031 (2000).

\bibitem{Barbosa:2002mg}
  A.~L.~Barbosa, L.~C.~T.~Guillen and J.~G.~Pereira,
  Phys.\ Rev.\ D {\bf 66}, 064028 (2002).

\bibitem{Fiorini:2013hva}
  F.~Fiorini, P.~A.~Gonzalez and Y.~Vasquez,
  arXiv:1304.1912 [gr-qc].

\bibitem{Bamba:2013fta}
  K.~Bamba, S.~'i.~Nojiri and S.~D.~Odintsov,
  Phys.\ Lett.\ B {\bf 725}, 368 (2013).

\bibitem{Geng:2014yya}
  C.~Q.~Geng, L.~W.~Luo, H.~H.~Tseng,
  Class.\ Quant.\ Grav.\  {\bf 31}, 185004 (2014) [arXiv:1403.3161 [hep-th]].

\bibitem{Einstein:teleaction}
  A.~Einstein,
  Sitzungsber. Preuss. Akad. Wiss., {\bf 1929}(X), 156–159, (1929).

\bibitem{Fujii:2003pa} 
  Y.~Fujii and K.~Maeda,
  ``The scalar-tensor theory of gravitation,''
  Cambridge, USA: Univ. Pr. (2003) 240 p

\bibitem{Carroll:2004st} 
  S.~M.~Carroll,
  ``Spacetime and geometry: An introduction to general relativity,''
  San Francisco, USA: Addison-Wesley (2004) 513 p

\end{thebibliography}
\end{document}